\documentclass[iop]{emulateapj}
\usepackage{epsfig}
\usepackage{apjfonts}
\usepackage{amsmath}

\begin{document}

\title{A double neutron star merger origin for the cosmological relativistic
fading source PTF11agg?}

\author{Xue-Feng Wu$^{1,2,*}$, He Gao$^{1,3}$, Xuan Ding$^{1}$, Bing Zhang$^{3,4,5}$, Zi-Gao Dai$^{6}$, Jian-Yan Wei$^{7}$}
\affil{$^1$Purple Mountain Observatory, Chinese Academy of Sciences, Nanjing, 210008, China; xfwu@pmo.ac.cn\\
  $^2$Chinese Center for Antarctic Astronomy, Chinese Academy of Sciences, Nanjing, 210008, China\\
  $^3$Department of Physics and Astronomy, University of Nevada Las Vegas, NV 89154, USA\\
  $^4$Department of Astronomy, Peking University, Beijing 100871, China\\
  $^5$Kavli Institute of Astronomy and Astrophysics, Peking University, Beijing 100871, China\\
  $^6$School of Astronomy and Space Science, Nanjing University, Nanjing 210093, China\\
  $^7$National Astronomical Observatories, Chinese Academy of Sciences, Beijing 100012, China\\}

\begin{abstract}
The Palomar Transient Factory (PTF)  team recently reported the discovery of a rapidly fading
optical transient source, PTF11agg. A long-lived scintillating radio
counterpart was identified, but the search for a high energy
counterpart showed negative results. The PTF team speculated that
PTF11agg may represent a new class of relativistic outbursts. Here
we suggest that a neutron star (NS)-NS merger system with a supra-massive magnetar
central engine could be a possible source to power such a transient,
if our line of sight is not on the jet axis direction of the system.
These systems are also top candidates for gravitational wave sources
to be detected in the advanced LIGO/Virgo era. We find that the PTF11agg
data could be explained well with such a model, suggesting that
at least some gravitational wave bursts due to NS-NS mergers
may be associated with such a
bright electromagnetic counterpart without a $\gamma$-ray trigger.
\end{abstract}

\keywords{gravitational waves, radiation mechanisms: non-thermal, stars: neutron}

\section{Introduction}

Recently, \cite{cenko13} (hereafter C13) reported a new discovery
from the Palomar Transient Factory (PTF), named as PTF11agg.
A multi-wavelength counterpart search was performed for this
rapidly fading optical transient. While a year-long scintillating
radio counterpart was identified, no high energy counterpart was
found. Based on a late-time, deep optical observation which revealed
a faint, quiescent source at the transient location, C13 suggested a
cosmological origin for this transient. If so, a relativistic ejecta
is required in order to explain the incoherent radio emission.

Many cosmological sources have been known (or proposed) to be able
to generate relativistic ejecta, such as active galactic nuclei
( Ghisellini et al. 1993; Krawczynski \& Treister 2013),
gamma-ray bursts (GRBs; Zhang \& M{\'e}sz{\'a}ros 2004;
M{\'e}sz{\'a}ros 2006; Gehrels \& Razzaque 2013), tidal disruption
of a star by a supermassive black hole
\citep{burrows11,bloom11,cenko12,lei11,lei13}, as well as
core-collapse supernova without GRB association \citep{soderberg10}.
Among these sources, the GRB afterglows have the most similar
observed properties to PTF11agg, regardless of the fact that
PTF11agg has no high energy counterparts. If PTF11agg is indeed a
GRB afterglow, the lack of a high energy signature could be explained
in two ways: an on-axis burst but without a high energy band trigger
due to a lack of satellite coverage; or an off-axis burst that gives
rise to an ``orphan" afterglow emerging due to a viewing-angle effect
\citep{rhoads97,nakar02}. C13 considered both possibilities: it
turns out the likelihood of an ``untriggered" on-axis long GRB being
discovered by PTF is quite small ($¡Ö2.6\%$), and the off-axis
afterglow model fails to interpret both the optical and radio data. They
therefore speculated that PTF11agg may represent a new class of
relativistic outburst.

Numerical simulations show that binary neutron star (NS) mergers would
launch a mildly anisotropic outflow during the merger process,
which has a typical mass of about $10^{-4} \sim
10^{-2}M_{\odot}$ and a typical velocity of about $0.1-0.3 c$
(where $c$ is the speed of light)
\citep[e.g.][]{rezzolla11,rosswog13,hotokezaka13}. Moreover, if
the equation of state of nuclear matter is stiff as supported by the
current data, a stable supra-massive magnetar would form after the merger
\citep{dai06,zhang13,giacomazzo13}. If so, the proto-magnetar would
eject a near-isotropic Poynting flux dominated outflow.
Similar to the GRB pulsar energy injection model \citep{dai98a,dai98b,zhang01},
the magnetar would accelerate the ejecta to a mildly or even highly
relativistic speed,
producing a strong external shock upon interaction with the ambient
medium to give rise to strong broad-band emission \citep{gao13a}.
Hereafter, we define such a model as the ``double neutron star (DNS) merger''
afterglow model.
Such an afterglow signal is
similar to GRBs, which can naturally explain the lack of a
high-energy counterpart. We suggest that PTF11agg might be the
first recognized detection of MDNSM afterglow emission.

\section{Observations of PTF11agg}

PTF11agg was first detected by the Palomar 48\,inch Oschin
telescope at 5:17:11 on 2011 January 30, and is located at
R.A.(J2000.0) $= 08^{\mathrm{h}} 22^{\mathrm{m}}
17.195^{\mathrm{s}}$, decl.(J2000.0) $= +21^{\circ} 37\arcmin
38\farcs26$. In the $R$ band, the source shows decay behavior from the
very beginning, with $R = 18.26 \pm
0.05$\,mag in the first detection image and a faint last detection
$R = 22.15 \pm 0.33$ mag on 2011 February 1. Checking back to
2009 November, no optical emission was reported at this
location. Late-time ($\Delta t > 1$\,month) deep optical
observation revealed a faint, unresolved source in $g^{\prime}$ and $R$
bands at R.A.(J2000.0) $= 08^{\mathrm{h}} 22^{\mathrm{m}}
17.202^{\mathrm{s}}$, decl.(J2000.0) $= +21^{\circ} 37\arcmin
38\farcs26$. Based on this detection, C13 speculated that the
redshift of PTF11agg should fall somewhere in the range $0.5
\lesssim z \lesssim 3.0$. The $R$ band light curve of
PTF11agg could be fitted well by a power-law with a best-fit
index $\alpha = 1.66 \pm 0.35$, if $t_{0} = $23:34 UT ($\pm 1.7$\,hr)
on 2011 January 29 is taken as the onset time.

Besides optical observation, the Karl G.~Jansky Very Large Array
(\citealt{pcb+11}) was also employed to observe the radio counterpart
of PTF11agg, starting from 2011 March 11, with a total bandwidth
8\,GHz and local oscillator frequency 93.6\,GHz. The spectral
energy distribution  in the radio band was constructed at two epochs
on 2011 March 14 and 2011 April 7, and both can be fitted with a
power law with an index
$\beta = 1/3$ (convention $F_\nu \propto \nu^\beta$).
Based on the constraints from the angular diameter of the
emitting region, C13 inferred that PTF11agg was initially at least
modestly relativistic.

C13 also checked the archival data from three primary high-energy
facilities for GRB triggers, i.e. InterPlanetary Network \citep{hga+10}, the
Gamma-ray Burst Monitor on the \textit{Fermi} spacecraft
\citep{mlb+09}, and Burst Alert Telescope on the \textit{Swift}
spacecraft \citep{bbc+05}). No temporally coincident triggers were
reported in the direction of PTF11agg. The
X-Ray Telescope ( \citealt{bhn+05}) on \textit{Swift}
was also later employed to observe the location of PTF11agg on
2011 March 13, but no X-ray source was detected.

\section{DNS merger afterglow model and application to PTF11agg}

\cite{gao13a} recently proposed that the DNS merger scenario with a
supra-massive magnetar central engine could power bright, broad-band
electromagnetic signals, behaving similarly to GRB afterglow
emission. The basic picture of such a scenario is the following:

A stable supra-massive magnetar has a total spin
energy $E_{\rm{rot}}=(1/2)I \Omega_{0}^{2} \simeq 2\times
10^{52} I_{45} P_{0,-3}^{-2} ~{\rm erg}$ (with $I_{45} \sim 1.5$
for a massive NS), where $P_{0}$ is the initial spin
period of the magnetar. Throughout this letter, we use the convention
$Q=10^n Q_n$ in cgs units, except ejecta mass $M_{\rm ej}$, which is
in units of solar mass $M_{\odot}$.
The spindown
luminosity and the characteristic spindown time scale critically
depend on the dipole magnetic field strength $B_{p}$ and initial
spin period $P_0$ (which may be close to the break-up limit), i.e. $L_{\rm
sd} = L_{\rm sd,0}/(1+t/t_{\rm sd})^2$, where $L_{\rm sd,0} \simeq
10^{49} ~{\rm erg~s^{-1}}~B^{2}_{p,15}R_{6}^{6}P_{0,-3}^{-4}$, and
$t_{\rm{sd}} \simeq 2 \times 10^3 ~{\rm
s}~I_{45} B_{p,15}^{-2} P_{0,-3}^2 R_6^{-6}$, where $R=10^6R_6$ cm
is the stellar radius. The dynamics of the blastwave is defined by
energy conservation \citep{gao13a}
\begin{eqnarray}\label{Dyn}
L_{\rm{0}}t=(\gamma-1)M_{\rm ej}c^2+(\gamma^{2}-1)M_{\rm sw}c^2,
\end{eqnarray}
where $L_0=\xi L_{\rm sd,0}$ is the magnetar injection luminosity
into the blastwave, and $M_{\rm sw}=(4\pi/3)R^3nm_p$ is the swept-up
mass from the interstellar medium.

Initially, one has $(\gamma-1)M_{\rm ej}c^2\gg(\gamma^{2}-1)M_{\rm
sw}c^2$, so that the kinetic energy of the ejecta would increase
linearly with time until $t={\rm min}(t_{\rm sd}, t_{\rm dec})$,
where the deceleration time $t_{\rm{dec}}$ is defined by the
condition $(\gamma-1)M_{\rm ej}c^2=(\gamma^{2}-1)M_{\rm sw}c^2$.
After $t_{\rm{dec}}$, the blastwave enters a self-similar phase
described by the Blandford-McKee self-similar solution \citep{bm76}.
In this phase, the dynamics of the blastwave is only determined by a
few parameters (e.g. the total energy of the system and the ambient
density).

During all of the dynamical phases for the ejecta, a strong external
shock would be formed upon interaction with the ambient medium,
where particles are believed to be accelerated, giving rise to
broad-band synchrotron radiation.

Assuming the spectrum of accelerated electrons is a power-law
function with the index of $p$, and that a constant fraction
$\epsilon_e$ of the shock energy is distributed to electrons, we
could derive the minimum injected electron Lorentz factor as (for
$p>2$)
\begin{eqnarray}
\gamma_m=\frac{p-2}{p-1}\epsilon_e(\gamma -1) \frac {m_p} {m_e},
\end{eqnarray}
where  $m_p$ and $m_e$ are proton mass and electron mass
respectively. Also assuming that a constant fraction $\epsilon_B$ of
the shock energy density is distributed to the magnetic energy
density behind the shock, one can obtain
\begin{eqnarray}
B=(8 \pi e\epsilon_B)^{1/2},
\end{eqnarray}
where $e$ is the energy density in the shocked region.

For synchrotron radiation, the observed radiation power and the
characteristic frequency of an electron with Lorentz factor
$\gamma_e$ are given by \citep{rybicki79}
\begin{eqnarray}
\label{power} P(\gamma_e) \simeq \frac 4 3 \sigma_T c \gamma^2
\gamma_e^2 \frac {B^2} {8\pi},
\end{eqnarray}
\begin{eqnarray}
\label{freq} \nu(\gamma_e) \simeq \gamma \gamma_e^2 \frac {q_e B} {2
\pi m_e c},
\end{eqnarray}
where the factors of $\gamma^2$ and $\gamma$ are introduced to
transform the results from the frame of the shocked fluid to the
frame of the observer.

For an individual electron, the spectral power, $P_\nu$ (in unit of
${\rm erg\,Hz^{-1}\,s^{-1}}$) varies as $\nu^{1/3}$ for
$\nu<\nu(\gamma_e)$, and cuts off exponentially for
$\nu>\nu(\gamma_e)$ \citep{rybicki79}. The peak power occurs at
$\nu(\gamma_e)$ with an approximate value of
\begin{eqnarray}
\label{flux} P_{\nu, \rm max}\approx\frac{P(\gamma_e)}{\nu(\gamma_e)}=
\frac {m_e c^2 \sigma_T} {3 q_e} \gamma B~.
\end{eqnarray}

On the other hand, one could estimate the life time of a
relativistic electron with Lorentz factor $\gamma_e$ in the observer
frame as
\begin{eqnarray}
\tau(\gamma_e)=\frac{\gamma\gamma_e m_e c^2}{\frac 4 3 \sigma_T c
\gamma^2 \gamma_e^2 \frac {B^2} {8\pi}}=\frac{6\pi m_e
c}{\gamma\gamma_e \sigma_T B^2 }~.
\end{eqnarray}
By setting $\tau(\gamma_e)=t$, a critical electron Lorentz factor
$\gamma_c$ is defined,
\begin{eqnarray}
\label{cool} \gamma_c= \frac{6\pi m_e c}{\gamma \sigma_T B^2t},
\end{eqnarray}
where $t$ refers to the time in the observer frame. Above
$\gamma_c$, synchrotron radiation cooling becomes significant, so
that the electron distribution shape should be modified.

A third characteristic frequency $\nu_a$, i.e., the synchrotron
self-absorption frequency, could be defined by equating the
synchrotron flux and the flux of a blackbody, i.e.
\begin{eqnarray}
I_{\nu}^{\rm syn}(\nu_a)=I_{\nu}^{bb}(\nu_a)\simeq2kT\cdot\frac{\nu_a^2}{c^2}
\end{eqnarray}
where the blackbody temperature is
\begin{eqnarray}
kT \simeq \rm{max}[\gamma_{a},\rm{min}(\gamma_c,\gamma_m)]m_ec^2,
\end{eqnarray}
and $\gamma_{a}$ is the corresponding electron Lorentz factor of
$\nu_a$ for synchrotron radiation, i.e. $\gamma_a=(4\pi m_e
c\nu_a/3eB)^{1/2}$ \citep[e.g.][]{saripiran99,kobayashizhang03a}.

In the Blandford-McKee regime, the characteristic synchrotron frequencies
and the peak synchrotron flux density $F_{\rm{\nu, \rm max}}=4\pi/3R^3n
P_{\nu, \rm max}$ for a constant circum-medium density could be expressed
 \citep{sari98,gao13b}.

\begin{eqnarray}
&&\nu_m=   8.1\times10^{11}~{\rm Hz}~(1+z)^{1/2}\left(\frac{p-2}{p-1}\right)^2E_{52}^{1/2}\epsilon_{e,-1}^{2}\epsilon_{B,-2}^{1/2}t_{5}^{-3/2}           ,\nonumber\\
&&\nu_c=    2.9\times10^{16}~{\rm Hz}~(1+z)^{-1/2}E_{52}^{-1/2}n_{0,0}^{-1}\epsilon_{B,-2}^{-3/2}t_{5}^{-1/2}           \nonumber\\
&&F_{\rm{\nu, \rm max}}=     1.1\times10^{4}~\mu {\rm
Jy}~(1+z) E_{52}^{}n_{0,0}^{1/2}\epsilon_{B,-2}^{1/2}D_{28}^{-2}          ,\nonumber\\
&&\nu_a=3.1\times10^{9}~{\rm
Hz}~(1+z)^{-6/5}\frac{g(p)}{g(3.2)}E_{52}^{1/5}n_{0,0}^{3/5}\epsilon_{e,-1}^{-1}\epsilon_{B,-2}^{1/5},\nonumber \\
\end{eqnarray}
where
$g(p)=\left(\frac{p-1}{p-2}\right)^{}(p+1)^{3/5}\left(\frac{\Gamma(\frac{3p+22}{12})\Gamma(\frac{3p+2}{12})}{\Gamma(\frac{3p+19}{12})\Gamma(\frac{3p-1}{12})}\right)^{3/5}$
is a numerical constant relate to $p$.

\begin{figure}[t!]
  \plotone{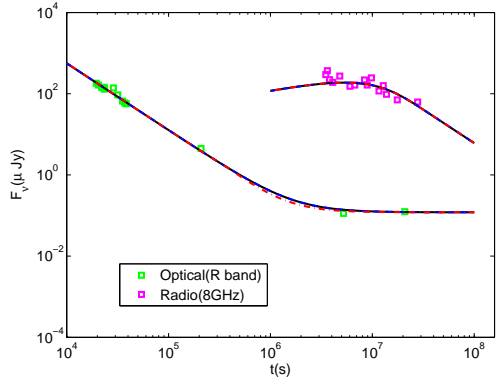}
  \caption{Optical and radio (8 GHz) light curves for PTF11agg, with best fittings by assuming different redshift for the source. The green square denotes
  the optical observation and the purple squares denote the radio data. The
  black solid line is for $z=0.5$, the blue dash line is for $z=1$ and the red dash-dot line is for $z=3$. }\label{figlc}
\end{figure}

Before applying the above DNS merger afterglow model to explain the
PTF11agg data, we first simply summarize the observational
properties of PTF11agg as follows \citep{cenko13}:

\begin{itemize}
\item Late time radio data suggest that the ejecta should be still relativistic
even at a very late epoch;
\item The optical light curve starts to decay at the very beginning of observation,
i.e., $t_{\rm s}=2\times10^4$s, with a simple power law decay slope
$\alpha=1.66\pm 0.35$. The first optical flux in $R$ band is about
$180~\rm \mu Jy$ (see Figure \ref{figlc});
\item The radio
band (8 GHz) light curve reached its peak around $10^7$ s, where the
peak flux is about $200 ~\rm \mu Jy$. The spectral slope for the
early radio spectral regime is about $\beta = 1/3$  (until
$5.8\times10^6$ s), implying that the radio peak should correspond
to $\nu_m$ crossing.
\end{itemize}

First, to reach a relativistic speed for the ejecta, we need
\begin{eqnarray}
\label{mcr} M_{\rm ej} \leq M_{\rm ej,c}
\end{eqnarray}
where $M_{\rm ej,c}\sim 6\times 10^{-3} M_\odot I_{45} P_{0,-3}^{-2}
\xi$ (defined by setting $E_{\rm rot}\xi = 2 (\gamma-1) M_{\rm
ej,c,2} c^2$), above which the blast wave would never reach a
relativistic speed \citep{gao13a}. Nevertheless, there is no obvious
break for late radio light curve, implying that the ejecta is still
in the relativistic regime (e.g. $\gamma-1 \geq 1$) until the end of the
observation $t_{\rm e}=3\times10^7$s. Since it is in the Blandford-McKee
stage, the ejecta evolves as $\gamma\propto t^{-3/8}$, we thus
have
\begin{eqnarray}
\label{gammacr} \rm{max}(\gamma_{\rm sd}-1,\gamma_{\rm
dec}-1)\geq\left(\frac{t_{\rm e}}{t_{\rm s}}\right)^{3/8}.
\end{eqnarray}

Second, since the $R$-band light curve starts to decay from the
beginning of the observation, we thus have
\begin{eqnarray}
\label{tcr} {\rm max}(t_{\rm sd}, t_{\rm dec}) \leq t_{\rm s}.
\end{eqnarray}
On the other hand, since the radio light curve (see Figure
\ref{figlc}) implies that $\nu_m$ crosses the 8 GHz band at about
$5.8\times10^6~\sim~1.2\times10^7$, we thus have
\begin{eqnarray}
\label{numcross}
8.1\times10^{11}(1+z)^{1/2}\left(\frac{p-2}{p-1}\right)^2E_{52}^{1/2}\epsilon_{e,-1}^{2}
\tilde{\epsilon}_{B,-2}^{1/2}\left(\frac{t_{\rm cross}}{10^5}\right)^{-3/2}=8\times10^9, \nonumber\\
\end{eqnarray}
where
\begin{eqnarray}
\label{numcrosst} 5.8\times10^6<t_{\rm cross}<1.2\times10^7.
\end{eqnarray}

Moreover, from $\nu_m\propto t^{-3/2}$, one could easily estimate
that $\nu_m(t_{\rm s})<\nu_{\rm opt}=5\times10^{14}\rm Hz$. We thus
expect the optical band to fall into the spectrum regime $ \nu_m
< \nu_{\rm opt} < \nu_c$. Consequently the observed temporal decay
index ($\alpha_{\mathrm{opt}} = 1.66 \pm 0.35$) can be translated
directly into the electron spectral index $p$, i.e., for a
constant-density medium as suggested by C13, we find $p = 3.2 \pm
0.47$. At $t_{\rm s}$, we have
\begin{eqnarray}
\label{opt} f_{\nu}(t_{\rm
s})=F_{\nu, \rm max}\left(\frac{\nu}{\nu_m}\right)^{-1.1}\approx 180 \rm{\mu
Jy}.
\end{eqnarray}

Based on the peak flux in the radio band, we get
\begin{eqnarray}
\label{rpeak} 1.1\times10^{4}~(1+z)
E_{52}^{}n_{0,0}^{1/2}\tilde{\epsilon}_{B,-2}^{1/2}D_{28}^{-2}\approx200
\end{eqnarray}
Moreover, the self-absorption frequency should fall below the radio
frequency range, i.e.,
\begin{eqnarray}
\label{nua}
3.1\times10^{9}(1+z)^{-6/5}E_{52}^{1/5}n_{0,0}^{3/5}\epsilon_{e,-1}^{-1}\tilde{\epsilon}_{B,-2}^{1/5}<8\times10^9.
\end{eqnarray}

Note that the ejecta formed in a DNS merger system would
expand into a pulsar wind bubble created by the progenitor pulsars
\citep{gallant99}. The radius of the bubble could be about $10^{17}$ cm
\citep{konigl02}. The value of $\epsilon_B$ should be relatively
large in the bubble \citep{konigl02}. One can easily find that the
optical signals of PTF11agg are emitted within this bubble radius
(with $\epsilon_B$), while the radio emission is emitted from outside
(with $\tilde{\epsilon}_{B}=\eta\epsilon_{B}$, and $0<\eta<1$, see
equations \ref{numcross}, \ref{rpeak} and \ref{nua}).

From equations \ref{mcr}-\ref{tcr}, we get the following constraints on
the ejecta mass and spin down luminosity:

\begin{eqnarray}
&&M_{\rm ej}\leq 1.1\times 10^{-3}\xi M_{\odot}\nonumber\\
&&L_{\rm{sd,0}}\geq 1.5\times10^{48}\xi~\rm{erg~s^{-1}}~.\nonumber\\
\end{eqnarray}

Combining equations \ref{numcross}-\ref{nua}, we have

\begin{eqnarray}
&&\eta^{1/2}=0.006\xi^{-1}(1+z)^{-1}n^{-1/2}\epsilon_{B,-2}^{-1/2}D_{28}^{2}\nonumber\\
&&n^{1/2}=0.93\xi^{-1.55}(1+z)^{-1.55}\epsilon_{e,-1}^{-2.2}\epsilon_{B,-2}^{-1.05}D_{28}^{2}\nonumber\\
&&1240(1+z)^{-1.05}\xi^{-1.05}~\leq \epsilon_{e,-1}^{4.2}\epsilon_{B,-2}^{1.05}~\leq3921(1+z)^{-1.05}\xi^{-1.05}~.\nonumber\\
\end{eqnarray}

For a given redshift, we only get three independent constraints on five
unknown parameters, i.e., $\xi$, $n$, $\epsilon_e$,
$\epsilon_B$ and $\eta$ (note that $p=3.2$ is fixed). This
leaves us some degeneracy in choosing parameters to fit the data.

In the following, we fix $\xi=1/3$ (i.e. $\sim 10^{52}
\rm erg$ spin down energy injected into the ejecta) and
$\epsilon_e=0.4$, and then fit the optical and radio data for
different redshifts by adopting appropriate values for $\epsilon_B$,
$n$ and $\eta$. One fitting result is shown in Figure \ref{figlc},
and the adopted parameters are collected in Table 1. With the
chosen parameters, we have derived the corresponding X-ray flux,
which is found to be consistent with the non-detection limitation.

In view of the fact that PTF11agg might be the first recognized
candidate for DNS merger afterglow, it would be helpful to compare
the inferred shock parameter values with those of other relativistic
shock related phenomena, such as GRBs. Recently, \cite{santana13}
performed a careful literature search for $\epsilon_e$ and
$\epsilon_B$, and found that $\epsilon_e \sim 0.02-0.6$ and
$\epsilon_B \sim 3.5\times10^{-5}-0.33$ were favored by the
observations.
With GRB optical afterglow data, \cite{liang13} found that the
electron spectral index $p$ is distributed in the range from 2 to
3.5\footnote{Note that medium density profile is relevant for
determining the electron index $p$. For a general circumburst medium density
profile $n\propto r^{-k}$ as adopted in \cite{liang13}, the derived $p$ would be 
somewhat larger than the value derived by assuming a constant density of the medium}. 
Moreover, Shen et al. (2006) performed a general investigation
for relativistic sources, such as GRBs (with both prompt and afterglow data), 
blazars and pulsar wind nebulae, and found a similar broad distribution of $p$. 
Our inferred parameters from the DNS merger afterglow model fall well within 
the ranges of these parameter distributions.

\begin{table}
\centering \caption{Adopt parameters for fitting the optical and
radio data of PTF11agg for different redshift.}
\begin{tabular}{lllllll}
\hline\hline

$z$  ~~~~~~    & ~~~~~~ $n~(\rm{cm^{-3}})$                    &  ~~~~~~$\epsilon_e$   & ~~~~~~$\epsilon_B$   & ~~~~~~$\eta$ & ~~~~~~$\xi$&~~~~~~$p$\\
\hline
$0.5$ ~~~~~~   & ~~~~~~ $1.0\times10^{-4}$     &  ~~~~~~$0.4$          & ~~~~~~$0.08$         & ~~~~~~$0.09$ &~~~~~~ $0.3$&~~~~~~$3.2$\\
$1$   ~~~~~~   & ~~~~~~ $2.4\times10^{-3}$     &  ~~~~~~$0.4$          & ~~~~~~$0.06$         & ~~~~~~$0.09$ &~~~~~~ $0.3$&~~~~~~$3.2$\\
$3$   ~~~~~~   & ~~~~~~ $0.26$                 &  ~~~~~~$0.4$          & ~~~~~~$0.03$         & ~~~~~~$0.09$ &~~~~~~ $0.3$&~~~~~~$3.2$\\
\hline
\end{tabular}
\end{table}

\section{Discussion}

We have proposed a DNS merger origin for the cosmological relativistic
fading source PTF11agg. Based on the observational properties of
PTF11agg, we analytically constrained the parameter space for the DNS
merger afterglow model and then fit the multi-band data
by adopting appropriate parameter values. We find that the DNS
merger afterglow model could fit both the optical and radio data well
regardless of the source redshift. If our interpretation is correct,
the following implications can be inferred:

First, the next generation gravitational-wave (GW)
detectors are expected to detect GW signals from mergers of two
compact objects, with DNS mergers as primary targets.
PTF11agg-like
transients could be potential electromagnetic counterparts to such
GW signals. The study of PTF11agg-like transients  would not only
shed light on the nature of the DNS merger scenario itself, but also
contribute to identifying the astrophysical origin of GW signals.

Second, since DNS mergers are proposed to be the progenitor
of short GRBs, the lack of a high energy counterpart for PTF11agg
could be due to the fact that our line of sight is not along the
direction of the jet axis. If so, there exists the possibility of
simultaneously detecting a short GRB afterglow and (off-beam) PTF11agg-like
emission, when our line of sight is within the jet opening angle.
Since the PTF11agg-like emission component is Doppler de-boosted
with respect to the on-beam calculations \citep{gao13a}, it
is detectable only under favorable condition. Some short GRB
afterglow features may be accounted for within this picture
(H. Gao et al. 2013, in preparation).

Third, besides the bright afterglow discussed by \cite{gao13a}
and this Letter, the DNS merger scenario with a supra-massive central
engine was also proposed to give rise to other bright electromagnetic
counterparts: the dissipation of a
Poynting flux dominated outflow from proto-magnetar would power a
bright early X-ray afterglow \citep{zhang13}; the
magnetar wind would add energy to the ejecta, giving rise to a
brighter ``merger-nova'' than the pure r-process powered one
\citep{yu13,metzger13};
Recently the short GRB 130603B has attracted
attention by showing an infrared excess in its late emission
\citep{tanvir13,berger13}. Even though it was suggested
that this emission is consistent with the $r$-process powered
``kilonova" emission with a black hole central engine, some
authors have already pointed out that both the mergernova and
short GRB afterglow of this burst can be understood within the
scenario of a supra-massive magnetar central engine, as long
as a large fraction of magnetar spin-energy is lost, possibly
by GW radiation
\citep{fan13,metzger13}. If so,
the short-lived transient
emission from GRB 130603B and PTF11agg may be different manifestations
of the same intrinsic phenomenon with different viewing angles
and/or magnetar parameters.

Finally, due to the large uncertainty of the DNS merger event
rate and the fraction of mergers that produce stable magnetars, it is
difficult to predict the detection rate by blind surveys for
DNS merger afterglows \citep[e.g.][for a discussion]{yu13}.
C13 suggested the event rate of PTF11agg-like
sources is $\sim$5 times that of normal GRBs, namely $5~{\rm
Gpc^{-3}~yr^{-1}}$.
Assuming a
fraction $f$ of DNS mergers may give rise to PTF11agg-like events,
one can estimate the DNS merger events rate as $\dot{\mathcal{N}}
\sim 5/f~{\rm Gpc^{-3}~yr^{-1}}=50f^{-1}_{-1}~{\rm
Gpc^{-3}~yr^{-1}}$\citep{gao13c}, which is consistent with
predictions using other methods
\citep{phinney91,kalogera04,abadie10}.

We note that \cite{wang13} also proposed a different model
for PTF11agg within the same framework as this Letter. They assumed
that the magnetar wind injection is in the form of
electron/positron pairs rather than a Poynting flux, and they
interpreted the observed emission from the reverse shock region.
They obtained different model parameter values and different
spectral properties from ours. Future observations with a
larger sample of PTF11agg-like transients may be helpful to
distinguish between these two models, and consequently lead to a
diagnosis of the composition of the magnetar wind.

We acknowledge the National Basic Research Program (``973" Program)
of China under Grant No. 2014CB845800, 2013CB834900 and
2009CB824800. This work is also supported by the National Natural
Science Foundation of China (grant No. 11033002 \& 10921063).
XFW acknowledges support by the One-Hundred-Talents
Program and the Youth Innovation Promotion Association of Chinese
Academy of Sciences.

\end{document}